# Higher-order symmetry energy of nuclear matter and the inner edge of neutron star crusts

W. M. Seif [1] and D. N. Basu [2]

[1] Cairo University, Faculty of Science, Department of Physics, Giza 12613, Egypt
[2] Variable Energy Cyclotron Centre, 1/AF Bidhan Nagar, Kolkata 700 064, India

The parabolic approximation to the equation of state of the isospin asymmetric nuclear matter (ANM) is widely used in the literature to make predictions for the nuclear structure and the neutron star properties. Based on the realistic M3Y-Paris and M3Y-Reid nucleon-nucleon interactions, we investigate the effects of the higher-order symmetry energy on the proton fraction in neutron stars and the location of the inner edge of their crusts and their core-crust transition density and pressure, thermodynamically. Analytical expressions for different-order symmetry energy coefficients of ANM are derived using the realistic interactions mentioned above. It is found that the higher-order terms of the symmetry energy coefficients up to its eighth-order ($E_{sym\,8}$) contributes substantially to the proton fraction in $\beta$ stable neutron star matter at different nuclear matter densities, the core-crust transition density and pressure. Even by considering the symmetry energy coefficients up to $E_{sym\,8}$, we obtain a significant change of about 40% in the transition pressure value from the one based on the exact equation of state. Using equations of state based on both Paris and Reid effective interactions which provide saturation incompressibility of symmetric nuclear matter in the range of $220\,MeV \leq K_0 \leq 270\,MeV$, we estimate the ranges $0.090\,fm^{-3} \leq \rho_t \leq 0.095\,fm^{-3}$ and $0.49\,MeV\,fm^{-3} \leq P_t \leq 0.59\,MeV\,fm^{-3}$ for the liquid core-solid crust transition density and pressure, respectively. The corresponding range of the proton fraction at this transition density range is found to be $0.029 \leq x_{p(t)} \leq 0.032$.

## I. INTRODUCTION

To understand many astrophysical phenomena and processes, we need to know accurate information about the density dependence of the isospin asymmetric nuclear matter (ANM) equation of state (EOS) and its isospin dependence as well [1], which are still largely unknown. Of course, the ANM equation of state plays a significant role in determining the different properties of neutron stars (NS) such as the proton fraction ($x_p$) in their matter and the critical density for the direct URCA process and consequently the cooling rate of NS. Also, the location of the inner edge of the NS crusts, their core-crust transition density ($\rho_t$) and transition pressure ($P_t$), and the crustal fraction of their moment of inertia as well as the critical frequency of the rotating NS are examples of such properties. Furthermore, the essential role of the equation of state of ANM in studying different aspects of nuclear structure such as the stability of neutron-rich nuclei [2] and its neutron skin thickness [3], surface energy and surface diffuseness of finite nuclei, the heavy-ion (HI) reaction dynamics and isospin diffusion is already established.

In terms of its different properties such as the energy per nucleon and incompressibility, the expansion of the ANM equation of state with respect to its density ($\rho$) and isospin asymmetry ($I$) is commonly used to study the nuclear matter, nuclear structure [1,4,5] and the NS properties [6,7,8]. For example, based on the M3Y-Paris and M3Y-Reid interactions and within a non-relativistic Hartree-Fock scheme [9], the two dimensional expansion of the ANM equation of state with respect to both $\rho$ and $I$, up to the eighth orders, is investigated in Ref. [4]. The M3Y-Paris [10] and M3Y-Reid [11] nucleon-nucleon (NN) effective interactions are derived from the bare NN interaction fitting of the G matrix. These realistic finite-range interactions are widely used in the nuclear reactions studies. However, it is found that the expansion of the EOS up to the quadratic term only, the parabolic approximation, expresses exactly $E_A(\rho, I)$ up to ANM density of $5\rho_0$ with $I \leq 0.75$ [4]. The fourth-order symmetry energy, $E_{sym\,4}(\rho)$, is needed to express the energy of pure neutron matter (PNM) at the higher densities above $4\rho_0$.

Within the framework of a relativistic mean-field formalism and based on the FSUGold and IU-FSU interactions, it is also shown that $E_{sym\,4}(\rho)$ may become non-negligible at high densities [6]. It is shown that the $E_{sym\,4}(\rho)$ contribution enhances the calculated proton fraction in $\beta$-stable $npe\mu$ matter at high densities, compared with the results based on the parabolic approximation of the EOS, and reduces the calculated core-crust transition density and transition pressure in NS [6]. Also, the ANM saturation density, energy per nucleon, and incompressibility as functions of $I$ (up to $I^8$) are linked in derived analytical formulas to the CDM3Y-K density dependent forms [12,13] of Paris and Reid interactions through 24 well defined characteristic quantities. These quantities represent the different order derivatives of the energy with respect to $\rho$ and $I$. It is shown that only 18 characteristic quantities are necessary to describe well the different saturation properties of ANM. These quantities include the symmetry energy coefficients up to $E_{sym\,8}$ and their derivatives with respect to density up to the fourth-order derivative of $E_{sym}$ ($E_{sym\,2}$), at the SNM saturation density point ($\rho_0$), as well as the quantities up to the fifth-order derivative of the SNM energy with respect to $\rho$, at $\rho_0$ [4]. It is seen that the quantities up to the fourth order terms give a good description of the saturation properties for ANM with $I \leq 0.3$ [4]. On the other hand, constraints on the incompressibility of ANM can be extracted experimentally by analyzing the isotopic dependence of the giant monopole resonance of neutron-rich nuclei [14,15] and the isospin diffusion data [16]. The constraints on the symmetry incompressibility either it refers to the $I^2$ coefficient in presence of higher order contributions or to the quadratic term as the only asymmetry term (parabolic approximation), as used in the most analysis, give some discrepancies among the different studies [4,5,17,18]. Consequently, the conclusion drawn from the different studies is that the widely used empirical parabolic approximation of the ANM equation of state may produce significant non-negligible errors in the calculated ANM properties and the different nuclear quantities related to them. This is clearly seen in particular under the extreme density and isospin asymmetry conditions. It is worth mentioning that the coefficients up to the $I^4$ dependence, $K_{04}$, or even up to the $I^6$, $K_{06}$, are required to reproduce the exact results of the $I$ dependence of the ANM saturation incompressibility reasonably up to the PNM for the calculations based on Reid interaction [4]. On the other hand, the quadratic term for the case of Paris interaction is found to be sufficient to express the exact results. This indicates that the influences of the higher-order contributions of the isospin asymmetry expansion of the ANM equation of state on the different nuclear properties are often model dependent.

In the present work, analytical formulas for different properties of ANM related to the neutron stars are derived based on the realistic M3Y-Paris and M3Y-Reid NN interactions in their CDM3Y-K density dependent forms. The effect of the higher-order symmetry energy $E_{sym\,n}(\rho)$ ($n = 2,4,6,8$) contribtions in determining different neutron stars properties is then investigated using the ANM equation of state based on the mentioned interactions. The outline of the present paper is the following. In the next section, the scheme used to derive the analytical formulas which express the different mentioned properties of ANM and neutron stars is outlined. The results based on equations of state characterized by SNM incompressibility values lying in the range of 220-250 MeV are presented and discussed in Sec. III. Also, comparisons of the present results with those obtained in recent studies based on different interactions are presented in the same section. Finally, section IV gives a brief summary and conclusions.

## II. THEORETICAL FRAMEWORK

### i. EOS of isospin asymmetric nuclear matter and the symmetry energy based on the M3Y effective interactions

In the frame work of a non-relativistic Hartree-Fock scheme [9], the ANM energy per nucleon based on the M3Y-Paris and M3Y-Reid NN interactions in their CDM3Y density dependent forms, is given as [13],

$$E_A(\rho, I) = \frac{3\hbar^2 k_F^2[(1+I)^{5/3} + (1-I)^{5/3}]}{20m} + f(\rho)\frac{\rho}{2}\left\{C_0 J_{00}^D + I^2 C_1 J_{01}^D + \frac{1}{4}\int [C_0 v_{00}^{Ex} B_0^2 + C_1 v_{01}^{Ex} B_1^2]d\vec{r}\right\},$$

$$B_0(I,r) = (1+I)\hat{j}_1(k_{Fn}r) + (1-I)\hat{j}_1(k_{Fp}r),$$

$$B_1(I,r) = (1+I)\hat{j}_1(k_{Fn}r) - (1-I)\hat{j}_1(k_{Fp}r). \tag{1}$$

Here, the first term represents the kinetic energy and $I$ is the isospin asymmetry parameter, $I = (\rho_n - \rho_p)/\rho$. $k_{Fn}, k_{Fp}$ and $k_F$ denote the neutron, proton and total Fermi momenta, respectively. $m(MeV/c^2)$ is the nucleonic mass. $v_{00}^{D(Ex)}$ and $v_{01}^{D(Ex)}$ are the central isoscalar and isovector direct (exchange) components [4,9,19] of the M3Y-Paris [10] and M3Y-Reid [11] NN interactions, respectively. In terms of the central direct components, $J_{00}^D$ and $J_{01}^D$ in Eq. (1) are given as $J_{00}^D = \int v_{00}^D(r)d\vec{r}$ and $J_{01}^D = \int v_{01}^D(r)d\vec{r}$. Also, $\hat{j}_1(x)$ is defined in terms of the first order spherical Bessel function as $\hat{j}_1(x) = 3j_1(x)/x$.

To reproduce the saturation properties of cold nuclear matter (NM), the original M3Y effective interaction has been modified by a phenomenological density dependent factor, $F(\rho)$, as [12]

$$v_{00(01)}^{D(Ex)}(\rho, r) = F_{0(1)}(\rho) v_{00(01)}^{D(Ex)}(r)$$

$$F_{0(1)}(\rho) = C_{0(1)} f(\rho) = C_{0(1)}(1 + \alpha e^{-\beta\rho} - \gamma\rho).$$

Here, the density dependent factors of both the isoscalar ($v_{00}^{D(Ex)}(\rho, r)$) and the isovector ($v_{01}^{D(Ex)}(\rho, r)$) parts the NN interaction are assumed to be of the same form but with the different $C_0$ and $C_1$ strengths, respectively [9,20]. The parameters of the CDM3Y density dependent factors of both the M3Y-Paris and M3Y-Reid interactions are linked to the saturation properties of SNM in simple derived relations [13]. These relations can be used to get the density dependence parameters which generate any EOS characterized with a proposed SNM saturation incompressibility value.

Taking accounts of the protons-neutrons exchange symmetry considerations concerning the charge symmetry of nuclear forces, the ANM equation of state in terms of the energy per nucleon ($E_A(\rho, I)$) can be expanded around $I = 0$ as

$$E_A(\rho, I) = E_A(\rho, 0) + E_{sym}(\rho)I^2 + E_{sym\,4}(\rho)I^4 + E_{sym\,6}(\rho)I^6 + E_{sym\,8}(\rho)I^8 + \cdots. \quad (2)$$

The notation $E_{sym}(\rho)$, as traditionally used in literature, is used here for the leading coefficient of the quadratic term $I^2$ instead of a systematic notation $E_{sym\,2}$. According to Eqs. (1) and (2), the different symmetry energy coefficients are given as [4],

$$\begin{aligned}
E_{sym}(\rho) &= \frac{\hbar^2 k_F^2}{6m} + \frac{f(\rho)\rho}{2}\left\{C_1 J_{01}^D + \frac{1}{4}\int\left[C_0 v_{00}^{Ex} B_{00} B_{00}^{(2)} + C_1 v_{01}^{Ex}\left(B_{10}^{(1)}\right)^2\right]d\vec{r}\right\} \\
E_{sym\,4}(\rho) &= \frac{\hbar^2 k_F^2}{162m} + \frac{f(\rho)\rho}{96}\int[C_0 v_{00}^{Ex} M_0 + C_1 v_{01}^{Ex} M_1]d\vec{r} \\
E_{sym\,6}(\rho) &= \frac{77\hbar^2 k_F^2}{20\,(3^7)m} + \frac{f(\rho)\rho}{6!\,4}\int[C_0 v_{00}^{Ex} M_2 + C_1 v_{01}^{Ex} M_3]d\vec{r} \\
E_{sym\,8}(\rho) &= \frac{1309\hbar^2 k_F^2}{80\,(3^9)\,m} + \frac{f(\rho)\rho}{8!\,4}\int[C_0 v_{00}^{Ex} M_4 + C_1 v_{01}^{Ex} M_5]d\vec{r},
\end{aligned} \quad (3)$$

where $M_0 = 3\left(B_{00}^{(2)}\right)^2 + B_{00} B_{00}^{(4)}$, $M_1 = 4B_{10}^{(1)} B_{10}^{(3)}$, $M_2 = 15 B_{00}^{(2)} B_{00}^{(4)} + B_{00} B_{00}^{(6)}$, $M_3 = 10\left(B_{10}^{(3)}\right)^2 + 6B_{10}^{(1)} B_{10}^{(5)}$, $M_4 = 35\left(B_{00}^{(4)}\right)^2 + 28 B_{00}^{(2)} B_{00}^{(6)} + B_{00} B_{00}^{(8)}$, $M_5 = 56 B_{10}^{(3)} B_{10}^{(5)} + 8 B_{10}^{(1)} B_{10}^{(7)}$, $B_{00(10)} \equiv B_{0(1)}(I = 0)$, and $B_{00(10)}^{(n)} \equiv \left.\frac{\partial^n B_{0(1)}}{\partial I^n}\right|_{I=0}$.

## ii. The neutron stars properties based on the M3Y effective interactions.

In terms of the proton fraction $x = \rho_p/\rho$ in ANM, its isospin asymmetry becomes $I = 1 - 2x$. Thermodynamically, the chemical equilibrium of the direct URCA reactions, $n \to p + e^- + \tilde{v}_e$ and $p + e^- \to n + v_e$, in $\beta$-stable matter ($npe\mu$) of neutron stars yields [7],

$$\mu_e = \mu_n - \mu_p = 2\frac{\partial E_A(\rho, I)}{\partial I}. \quad (4)$$

$\mu_n$, $\mu_p$ and $\mu_e$ are the chemical potentials for the neutrons, protons and electrons, respectively. However, muons start to appear at $\rho \geq \rho_0$ [21] because they require a high chemical potential of electrons [22]. They provide then very little contribution to the chemical equilibrium. One should recall here that the core-crust transition density lies usually in the sub-saturation density [22]. Then, the charge neutrality condition of neutron stars implies $\rho_e = \rho_p = \rho x$ ($k_{F_e} = k_{F_p}$). In terms of the electron Fermi momentum, the chemical potential of the relativistic electrons becomes

$$\mu_e = \sqrt{k_{F_e}^2 c^2 + m_e^2 c^4} \approx k_{F_e} c = \hbar c (3\pi^2 \rho x)^{1/3}. \tag{5}$$

According to Eqs.(1), (4) and (5), the value of the proton fraction of $\beta$-stable matter ($x_p$) is determined by

$$x_p = \frac{1}{3\pi^2 \rho} \left( \frac{\hbar k_F^2}{2mc} \left[ (2-2x_p)^{\frac{2}{3}} - (2x_p)^{\frac{2}{3}} \right] \right.$$
$$\left. + \frac{2f(\rho)\rho}{\hbar c} \left\{ (1-2x_p) C_1 J_{01}^D + \frac{1}{4} \int \left[ C_0 v_{00}^{Ex} B_0 B_0^{(1)} + C_1 v_{01}^{Ex} B_1 B_1^{(1)} \right] d\vec{r} \right\} \right)^3. \tag{6}$$

Here, the coefficients $B_0$ and $B_1$, and their derivatives, $B_0^{(1)} = \frac{\partial B_0(I,r)}{\partial I}$ and $B_1^{(1)} = \frac{\partial B_1(I,r)}{\partial I}$ are calculated in terms of the isospin asymmetry of $\beta$-stable matter. Also, an approximate formula for the same quantity ($x_p$) can be obtained using the isospin asymmetry expansion of the ANM energy per nucleon, Eq.(2), in addition to Eqs. (4) and (5), as

$$x_p = \frac{1}{2} \left[ 1 - \frac{\hbar c (3\pi^2 \rho x_p)^{\frac{1}{3}}}{4 \left[ E_{sym} + 2E_{sym\,4}(1-2x_p)^2 + 3E_{sym\,6}(1-2x_p)^4 + 4E_{sym\,8}(1-2x_p)^6 \right]} \right]. \tag{7}$$

In terms of the pressure of baryons $(n,p)$, $P_N(\rho, x)$, [13] and that of electrons, $P_e(\rho, x)$ [8], we can derive the total pressure of the $npe$ matter using the expression of the energy per nucleon, Eq. (1), as

$$P(\rho, x) = P_N(\rho, x) + P_e(\rho, x)$$
$$= \frac{\hbar^2 k_F^2 \left[ (2-2x)^{\frac{5}{3}} + (2x)^{\frac{5}{3}} \right] \rho}{10m}$$
$$+ \frac{1}{2} \left( \rho^3 \acute{f}(\rho) + \rho^2 f(\rho) \right) \left\{ C_0 J_{00}^D + (1-2x)^2 C_1 J_{01}^D \right.$$
$$\left. + \frac{1}{4} \int (C_0 v_{00}^{Ex} B_0^2 + C_1 v_{01}^{Ex} B_1^2) d\vec{r} \right\} - \frac{\rho^2}{4} f(\rho) \int (C_0 v_{00}^{Ex} B_0 B_2 + C_1 v_{01}^{Ex} B_1 B_3) d\vec{r}$$
$$+ \frac{\hbar c}{12\pi^2} (3\pi^2 \rho x)^{4/3} \tag{8}$$

where $B_{\frac{2}{(3)}}(I,r) = (1+I) j_2(k_{Fn} r) \overset{+}{(-)} (1-I) j_2(k_{Fp} r)$. The prime on the density dependence function, $\acute{f}(\rho)$, denotes its derivative with respect to the nuclear density. The last term in Eq. (8) represents the electronic pressure contribution, assuming non-interacting Fermi gas of electrons [8]. Correspondingly, based on the isospin asymmetry expansion of $E_A(\rho, I)$, Eq. (2), the total pressure can be approximated as

$$P(\rho, x) = \rho^2 \left[ E_A^{[1]}(\rho, x=0.5) + E_{sym}^{[1]}(1-2x)^2 + E_{sym4}^{[1]}(1-2x)^4 + E_{sym6}^{[1]}(1-2x)^6 \right.$$
$$\left. + E_{sym8}^{[1]}(1-2x)^8 \right] + \frac{\hbar c}{12\pi^2} (3\pi^2 \rho x)^{4/3}, \tag{9}$$

where $E_{symj}^{[n]} (j = 0,2,4,6,8 \ \& \ n = 1,2,\dots) = \frac{d^n E_{symj}(\rho)}{d\rho^n}$.

Thermodynamically, the intrinsic stability conditions of a single phase for locally neutral matter under $\beta$ equilibrium can be mainly determined by the positivity of the compressibility of matter ($K_\mu$), under constant chemical potential [23],

$$K_\mu = -v^2 \left(\frac{\partial P}{\partial v}\right)_\mu = \left(\frac{\partial P}{\partial \rho}\right)_\mu = \frac{K(\rho, I)}{9} - \frac{\left(\frac{\partial^2 E_A(\rho, I)}{\partial \rho \partial I}\rho\right)^2}{\frac{\partial^2 E_A(\rho, I)}{\partial I^2}} > 0 \qquad (10)$$

Here, $v$ is the volume per baryon number and $\mu$ is the chemical potential. While $K(\rho, I) = 9\left[2\rho \frac{\partial E_A(\rho,I)}{\partial \rho} + \rho^2 \frac{\partial^2 E_A(\rho,I)}{\partial \rho^2}\right]$ is the known incompressibility coefficient of ANM ($np$ matter) [13], the last part in Eq.(10) arises from the partial derivative of the leptonic part of the pressure, with respect to $v$ ($\rho$). It is relevant to mention that there is another stability condition regarding the electrical capacitance of matter ($\chi_v = -(\partial q/\partial \mu)_v > 0$) but it is usually valid in our case [22,23]. However, the limiting baryon density that breaks these conditions will correspond to the core-crust (liquid-solid) phase transition in neutron stars. Using Eq. (1), we can find the explicit forms of the quantities mentioned in Eq. (10) as

$$K(\rho, I) = \frac{3\hbar^2 k_F^2 \left[(1+I)^{\frac{5}{3}} + (1-I)^{\frac{5}{3}}\right]}{2m}$$
$$+ 9\left(\frac{\rho^3}{2}\acute{f}(\rho) + 2\rho^2 \acute{f}(\rho) + \rho f(\rho)\right)\left\{C_0 J_{00}^D + I^2 C_1 J_{01}^D + \frac{1}{4}\int (C_0 v_{00}^{Ex} B_0^2 + C_1 v_{01}^{Ex} B_1^2) d\vec{r}\right\}$$
$$- \frac{9}{4}\left(2\rho^2 \acute{f}(\rho) + 3\rho f(\rho)\right)\int (C_0 v_{00}^{Ex} B_0 B_2 + C_1 v_{01}^{Ex} B_1 B_3) d\vec{r}$$
$$- \frac{9\rho}{4} f(\rho) \int \left(C_0 v_{00}^{Ex}(B_0 B_4 - B_2^2) + C_1 v_{01}^{Ex}(B_1 B_5 - B_3^2)\right) d\vec{r}, \qquad (10(a))$$

$$\frac{\partial^2 E_A(\rho, I)}{\partial \rho \partial I} = \frac{\hbar^2 k_F^2}{6m\rho}\left[(1+I)^{\frac{2}{3}} - (1-I)^{\frac{2}{3}}\right]$$
$$+ \left(\acute{f}(\rho)\rho + f(\rho)\right)\left\{I C_1 J_{01}^D + \frac{1}{4}\int \left[C_0 v_{00}^{Ex} B_0 B_0^{(1)} + C_1 v_{01}^{Ex} B_1 B_1^{(1)}\right] d\vec{r}\right\}$$
$$+ \frac{1}{4} f(\rho)\rho \int \left[C_0 v_{00}^{Ex}\left(B_0^{[1]} B_0^{(1)} + B_0 B_0^{(11)}\right)\right.$$
$$\left. + C_1 v_{01}^{Ex}\left(B_1^{[1]} B_1^{(1)} + B_1 B_1^{(11)}\right)\right] d\vec{r}, \qquad (10(b))$$

and

$$\frac{\partial^2 E_A(\rho, I)}{\partial I^2} = \frac{\hbar^2 k_F^2}{6m}\left[(1+I)^{-\frac{1}{3}} + (1-I)^{-\frac{1}{3}}\right]$$
$$+ f(\rho)\rho\left\{C_1 J_{01}^D\right.$$
$$\left. + \frac{1}{4}\int \left[C_0 v_{00}^{Ex}\left(\left(B_0^{(1)}\right)^2 + B_0 B_0^{(2)}\right) + C_1 v_{01}^{Ex}\left(\left(B_1^{(1)}\right)^2 + B_1 B_1^{(2)}\right)\right] d\vec{r}\right\}, \qquad (10(c))$$

where,

$$B_{4 \atop (5)}(I,r) = \{(1+I)[2j_2(k_{Fn}r) - (k_{Fn}r)j_3(k_{Fn}r)] \genfrac{}{}{0pt}{}{+}{(-)} (1-I)[2j_2(k_{Fp}r) - (k_{Fp}r)j_3(k_{Fp}r)]\}/3,$$

$$B_i^{(n)}(i=0,1,\ldots \& n=1,2,\ldots) \equiv \frac{\partial^n B_i}{\partial I^n}, \quad B_i^{[1]} = \frac{\partial B_i}{\partial \rho} \quad \text{and} \quad B_i^{(11)} = \frac{\partial^2 B_i}{\partial \rho \partial I}.$$

From the knowledge of the different coefficients of the symmetry energy, Eqs. (2) and (3), we can express the incompressibility condition (Eq. (10)) as

$$K_\mu = 2\rho \left[ E_A^{[1]}(\rho,0) + E_{sym}^{[1]} I^2 + E_{sym\,4}^{[1]} I^4 + E_{sym\,6}^{[1]} I^6 + E_{sym\,8}^{[1]} I^8 \right]$$
$$+ \rho^2 \left[ E_A^{[2]}(\rho,0) + E_{sym}^{[2]} I^2 + E_{sym\,4}^{[2]} I^4 + E_{sym\,6}^{[2]} I^6 + E_{sym\,8}^{[2]} I^8 \right]$$
$$- \frac{2I^2 \rho^2 \left[ E_{sym}^{[1]} + 2E_{sym\,4}^{[1]} I^2 + 3E_{sym\,6}^{[1]} I^4 + 4E_{sym\,8}^{[1]} I^6 \right]^2}{E_{sym} + 6E_{sym\,4} I^2 + 15E_{sym\,6} I^4 + 28E_{sym\,8} I^6} > 0, \qquad (11)$$

### III. CALCULATIONS AND DISCUSSION

In the following, the mentioned properties and quantities related to NS will be reviewed according to their dependence on the EOS and on the higher-order coefficients of symmetry energy. We shall investigate this using equations of state based on both Paris and Reid NN interactions in their CDM3Y density dependent forms.

Shown in Fig. 1 is the density dependence of the proton fraction ($x_p$) in $\beta$-stable $npe\mu$ ($npe$) matter. The calculations are based on the M3Y-Paris (Figs. 1(a) and 1(c)) and M3Y-Reid (Fig. 1(b)) NN interactions with different parameterizations [13] of their CDM3Y-K density dependent form. These parameterizations generate different equations of state characterized by saturation incompressibility values in the range of $220\ MeV \leq K_0 \leq 270\ MeV$. In Figs. 1(a) and 1(b), the density dependence of the proton fraction based on the different symmetry energy coefficients as predicted from the isospin asymmetry expansion of the ANM equation of state, Eq. (7), are compared with the exact calculations based on the full EOS, Eq. (6). The predicted proton fraction from the different mentioned equations of state based on Paris interaction are displayed as a function of density in Fig. 1(c). As can be seen in Figs. 1(a) and 1(b), the proton fraction based on both the M3Y-Paris and M3Y-Reid interactions show almost the same behavior with the ANM density. For the CDM3Y-240 ($K_0 = 240\ MeV$) density dependent form of Paris (Reid) interaction, it is found that the $\beta$-stable proton fraction increases with increasing the ANM density in the low density region reaching its maximum value, $x_p = 0.049$, at $\rho = 0.28\ fm^{-3}$ ($0.27\ fm^{-3}$). The proton fraction starts to decrease with increasing the density in the high density region, $\rho \geq 0.28\ fm^{-3}$. The $\beta$-stable NM becomes proton-free, and consequently electron-free, matter at $\rho = 1.08\ fm^{-3}$ ($0.95\ fm^{-3}$). Even by considering the higher-order symmetry energy coefficients, up to $E_{sym\,8}$, the exact maximum value of the proton fraction is not obtained. Compared to the exact value of the proton fraction ($x_p = 0.049$) based on the full EOS, the obtained maximum values of $x_p$ based on the isospin asymmetry expansion of the EOS up to $E_{sym}$ (parabolic approximation), $E_{sym\,4}$, $E_{sym\,6}$ and $E_{sym\,8}$ are 0.043(0.044), 0.043, 0.045 and 0.046, respectively. Regarding the effects of the EOS itself, the different density dependent parameterizations of the CDM3Y-Paris interaction that evolve equations of state with $220\ MeV \leq K_0 \leq 270\ MeV$ give a similar behavior for the proton fraction with density but with a slight shift in its maximum values, Fig. 1(c). The maximum values of $0.050 \geq x_p \geq 0.048$ are obtained in the range of $0.30\ fm^{-3} \geq \rho \geq 0.25\ fm^{-3}$. As seen in Fig. 1(c), the EOS does not affect the proton fraction in the $\beta$-stable $npe$ matter up to nuclear density of $\rho \approx 0.25\ fm^{-3}$. The main effect of the nuclear matter EOS appears in the density limit at which the $\beta$-stable ($npe$) NM becomes PNM, $x_p = 0$. We notice that this limiting density decreases as the stiffness of used EOS increases. The density range at which the $\beta$-stable $npe$ matter has non-zero proton fraction extends to $\rho = 1.54\ fm^{-3}$, $1.25\ fm^{-3}$, $1.08\ fm^{-3}$, $0.96\ fm^{-3}$ and $0.79\ fm^{-3}$ for the equations of state characterized with saturation incompressibility values of $K_0 = 220\ MeV, 230\ MeV, 240\ MeV, 250\ MeV$ and $270\ MeV$, respectively. As the obtained density dependence of the $\beta$-stable proton fraction yields a maximum value of $x_p = 0.049$, the direct URCA process in neutron stars would be then forbidden. This is because this process is only permitted when the proton fraction exceeds the critical value of $x_p = 1/9$ [24, 25]. This confirms the suggested relatively slow cooling process of NS $(n + (n,p) \to p + (n,p) + e^- + \tilde{\nu}_e$ and $p + (n,p) \to n + (n,p) + e^+ + \nu_e)$ instead of the fast one through direct nucleon URCA reactions [24]. This feature is consistent with the fact that there are no strong indications [26,27] that fast cooling occurs. It was also concluded theoretically that an acceptable EOS of ANM shall not allow the direct URCA process to occur in neutron stars with masses below 1.5 solar masses [28]. Even recent experimental observations that suggest high heat conductivity and enhanced core cooling process indicating the enhanced level of neutrino emission, were not attributed to the direct URCA process but were proposed to be due breaking and formation of neutron Cooper pairs [29-32].

Table I presents the core-crust transition density in neutron stars as extracted from the exact calculations of the incompressibility condition given by Eqs. (10, 10(a), 10(b) and 10(c)) using equations of state based on the CDM3Y-Paris and CDM3Y-Reid interactions. Also, the exact calculations of the transition pressure (Eq. (8)) and the corresponding proton fraction (Eq. (6)) are presented in the same table. The used equations of state are characterised with SNM saturation incompressibility range of $K_0 = 220\ \text{MeV} - 250\ \text{MeV}$. This range of $K_0$ is estimated for the nuclear matter EOS in various studies on the ground state properties of finite nuclei [5,16,33,34,35] and the nuclear reactions studies [12,20,36,37,38] based on different effective interactions including the M3Y-types. Moreover, only the CDM3Y-Paris (Reid) interaction characterized with SNM saturation incompressibility value of 240 MeV (230 MeV) predicts NM pressure which is inconsistent with previous constraints [39] on both SNM and PNM pressure [13]. Regarding the NS properties, we may question now the degree of accuracy of the calculations based on the isospin asymmetry expansion of the EOS given by Eq. (2). To this aim the approximate calculations of the transition density (Eq. (11)), pressure (Eq. (9)) and the corresponding proton fraction (Eq. (7)) using the different symmetry energy coefficients (Eq. (3)), and their derivatives with respect to density, are presented also in Table I. As shown in Table I, the core-crust transition density and pressure are estimated by the exact calculations based on the mentioned equations of state to be within the ranges of $0.090\ fm^{-3} \leq \rho_t \leq 0.095\ fm^{-3}$ and $0.49\ MeV\ fm^{-3} \leq P_t \leq 0.59\ MeV\ fm^{-3}$, respectively. The corresponding range of the equilibrium proton fraction value at this transition density range is obtained to be $0.029 \leq x_{p(t)} \leq 0.032$. However, the constraints obtained here on the core-crust transition density and pressure are consistent with those obtained in few similar studies [8,40] but they are somewhat higher than those obtained in other studies [6,8,22]. For example, using a modified Gogny (MDI) and 51 Skyrme interactions, within both dynamical and thermodynamical methods, the limits of $0.040\, fm^{-3} \leq \rho_t < 0.065\, fm^{-3}$ and $0.01\, MeV\, fm^{-3} < P_t \leq 0.26\, MeV\, fm^{-3}$ are imposed from constraints on the symmetry energy by the isospin diffusion data in HI collisions [22]. Also, the calculations performed in terms of the isospin asymmetry expansion of the EOS including the terms up to $E_{sym\,4}$ within a nonlinear relativistic mean field model based on the calibrated FSUGold and IU-FSU interactions yielded $0.051\, fm^{-3} \leq \rho_t \leq 0.077\, fm^{-3}$ and $0.24\, MeV\, fm^{-3} \leq P_t \leq 0.53\, MeV\, fm^{-3}$ [6]. Further, estimated constraints of $0.086\, fm^{-3} \leq \rho_t \leq 0.090\, fm^{-3}$ and $0.30\, MeV\, fm^{-3} \leq P_t \leq 0.76\, MeV\, fm^{-3}$ have been obtained in the framework of relativistic nuclear energy density functional, with adjusted functional to the binding energies of finite nuclei and their isovector properties [8]. As shown in Table I, while neglecting the higher-order symmetry energy coefficients increases both the calculated transition density and pressure with respect to their obtained exact values, it reduces slightly the calculated proton fraction. The change in the calculated transition pressure upon neglecting the higher-order coefficients of the symmetry energy is greater than that in the calculated transition density. A similar significant increase in the core-crust transition density and pressure due the parabolic approximation of the EOS have been demonstrated in recent studies based on nonrelativistic [22] and nonlinear relativistic mean field models [6]. As presented in Table I, the approximate calculations of the core-crust transition density in terms of the different symmetry energy coefficients up to the 2$^{nd}$ ($E_{sym}$), 4$^{th}$ ($E_{sym\,4}$), 6$^{th}$ ($E_{sym\,6}$) and 8$^{th}$ ($E_{sym\,8}$) orders lead to errors of about 14±2 %, 11±2 %, 7±2 % and 5±1 %, respectively, in comparison to the exact calculations. The errors in the corresponding values of the transition pressure (proton fraction) are $64 \pm 7\%$ ($7 \pm 3$ %), $53 \pm 5$ % ($6 \pm 3$ %), $44 \pm 3$ % ($3 \pm 3$ %) and $38 \pm 2$ % ($3 \pm 3$ %), respectively. It is interesting to note from the computed errors that disregarding the higher-order coefficients of the symmetry energy increases the errors in the obtained approximated values of the three mentioned quantities. In particular, we need to consider the symmetry energy coefficients up to the 8$^{th}$ order in the calculations to get the transition density and the corresponding proton fraction with an inevitable small error ($\leq 6\%$). On the other hand, even by considering the higher-order symmetry energy up $E_{sym\,8}$, the errors in the calculated transition pressure upon the isospin-asymmetry expansion of energy are always large ($\geq 40$ %). We notice also that the errors in the values of the mentioned quantities based on the isospin-asymmetry expansion of the EOS are generally smaller in the case of the M3Y-Paris interaction as compared to the M3Y-Reid one. This can be related to the different symmetry energy coefficients, based on the two interactions, and their slope parameters [8,22]. The symmetry energy coefficients and their corresponding slope parameters for the M3Y-Paris (M3Y-Reid) interaction, at the saturation density, are, $E_{sym}(\rho_0) = 30.85$ MeV (31.11 MeV), $E_{sym\,4}(\rho_0) = 0.10$ MeV ($-0.13$ MeV), $E_{sym\,6}(\rho_0) = 0.28$ MeV (0.31 MeV), $E_{sym\,8}(\rho_0) = 0.098$ MeV (0.102 MeV), $L(\rho_0) = 47.51$ MeV (50.98 MeV), $L_4(\rho_0) = -0.47$ MeV ($-1.43$ MeV), $L_6(\rho_0) = 0.58$ MeV (0.67 MeV) and $L_8(\rho_0) = 0.126$ MeV (0.133 MeV). These values are independent of the density dependence of the mentioned interactions and hence independent of the saturation incompressibility values [4]. However, a slight improvement in the calculations based the EOS expansion is achieved for the Paris interaction which yields smaller symmetry energy coefficients and slope parameters than those of the Reid interaction. We also observe a slight decrease in the transition pressure upon decreasing the symmetry energy coefficients. Actually, the criteria of accepting or rejecting a definite degree of accuracy for the approximate calculations of any of the proton fraction, the transition density or the transition pressure depends on how much the other physical predictions related to NS are sensitive to them.

## IV. SUMMARY AND CONCLUSION

Based on the realistic M3Y-Paris and M3Y-Reid effective interactions in their density dependent CDM3Y versions, the core-crust transition at the inner edge of neutron stars is investigated, thermodynamically. The intrinsic stability condition of the core-crust transition and consequently the transition density and pressure as well as the proton fraction of the $npe$ ($npe\mu$) matter under $\beta$ equilibrium are derived analytically based on exact EOS. Also, approximate analytical formulas for the same mentioned quantities are derived based the isospin asymmetry expansion of the ANM energy up to the $I^8$ coefficient of the symmetry energy. The different coefficients of the ANM symmetry energy as well as the NM pressure and compressibility are also derived analytically in the frame work of a non-relativistic Hartree-Fock scheme. However, based on equations of state characterized with SNM saturation incompressibility range of $K_0 = 220 \text{ MeV} - 270 \text{ MeV}$, the $\beta$-stable proton fraction of $npe$ matter is inferred to have its maximum value lying in the range of $0.25 \text{ } fm^{-3} \leq \rho \leq 0.3 \text{ } fm^{-3}$ and then it starts to decrease. The comparison of the approximate calculations with the exact ones shows that the higher-order symmetry energy coefficients up to $E_{\text{sym } 8}$ are needed to describe reasonably the proton fraction of the $\beta$ stable matter at the higher nuclear densities and also to obtain the core-crust transition density. The parabolic approximation of the EOS in terms of only the second order symmetry energy, do not affect seriously the proton fraction at the transition density. On the contrary, the calculations of the core-crust transition pressure upon the different symmetry energy coefficients up to $E_{\text{sym } 8}$ show a deviation of as high as 40 % from the exact calculations. The saturation incompressibility range of $K_0 = 220 \text{ MeV} - 250 \text{ MeV}$ which is estimated in different studies of ground state properties of finite nuclei and their reactions and from constraints on both the SNM and PNM pressure is employed to constrain the core-crust transition density and pressure. Using the exact calculations based on the CDM3Y-Paris and CDM3Y-Reid interactions, we estimate the constraints $0.090 \text{ } fm^{-3} \leq \rho_t \leq 0.095 \text{ } fm^{-3}$ and $0.485 \text{ } MeV \text{ } fm^{-3} \leq P_t \leq 0.589 \text{ } MeV \text{ } fm^{-3}$ on the core-crust transition density and pressure of neutron stars, respectively. The estimated constraint on the corresponding proton fraction at this density is obtained as $0.029 \leq x_{p(t)} \leq 0.032$.

**Fig. 1**: The density dependence of the proton fraction in the $\beta$-stable $npe$ matter based on the M3Y (a) Paris and (b) Reid NN interactions in their CDM3Y-240 density dependent form as extracted, exactly, from the calculations based on the full ANM equation of state, Eq. (6). The approximate calculations based on the expansion of the EOS up to different-order coefficients of the symmetry energy, $E_{sym\,n}(n = 2,4,6,8)$, are also presented for comparison. (c) shows the exact calculations based on different equations of state characterized by saturation incompressibility values in the range of $220\ MeV \leq K_0 \leq 270\ MeV$.

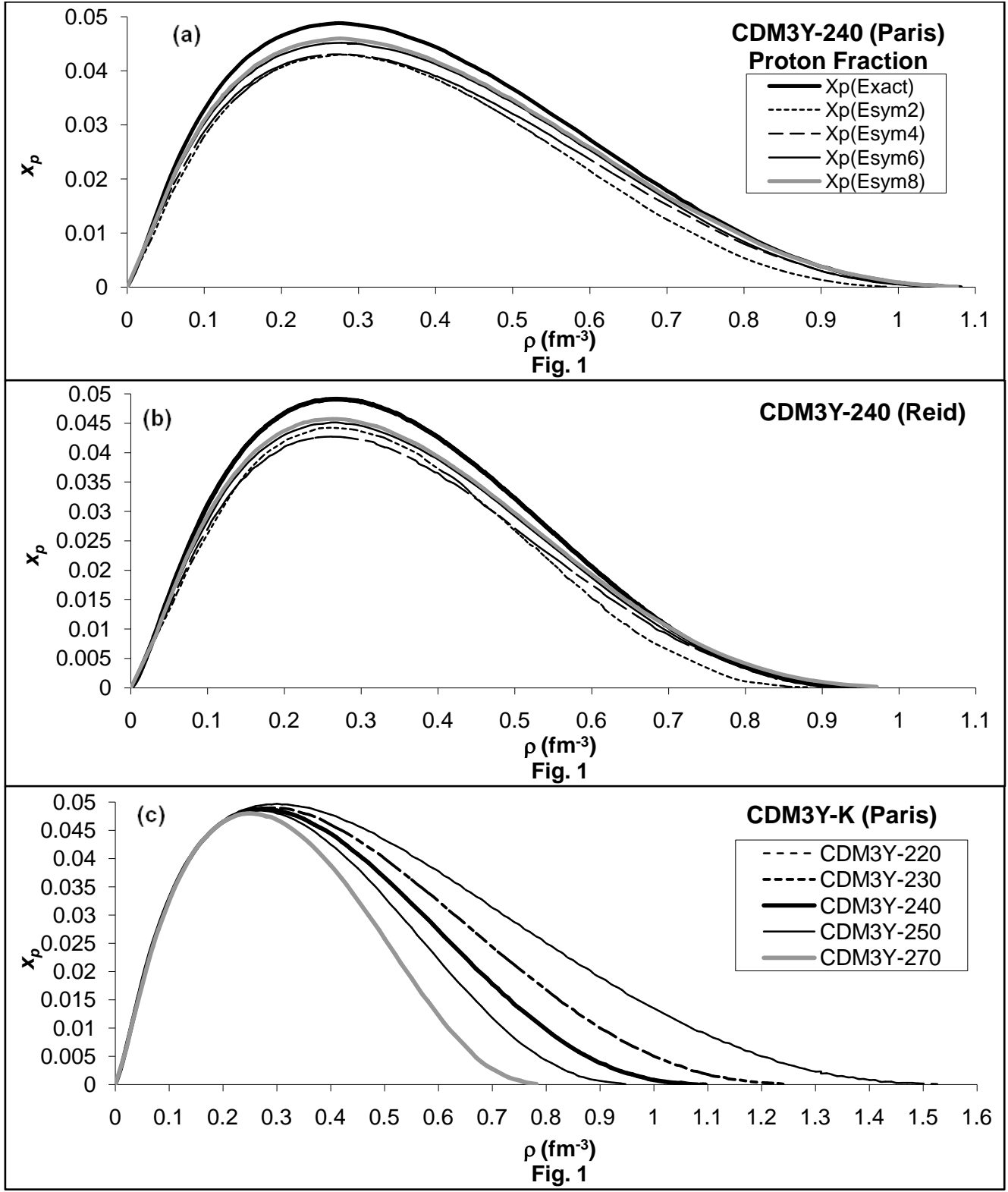

**Table I**: The exact calculations of the core-crust transition density, $\rho_t$ (Eqs. 10, 10(a), 10(b) and 10(c)), and pressure, $P_t$ (Eq. (8)) and the corresponding proton fraction $x_{p(t)}$ (Eq. (6)), in neutron stars as extracted exactly using full equations of state based on the CDM3Y-Paris and CDM3Y-Reid interactions. The calculations are performed using equations of state characterised with SNM saturation incompressibility range of $K_0 = 220$ MeV $- 250$ MeV. $\rho_t^{E_{symn}}$ (Eq. (11)), $P_t^{E_{symn}}$ (Eq. (9)), and $x_{p(t)}^{E_{symn}}$ (Eq. (7)) are the approximate values based on the isospin-asymmetry expansion of the EOS, up to different-order coefficients of the symmetry energy ($E_{sym\,n}(n = 2,4,6,8), E_{sym} \equiv E_{sym\,2}$).

| | M3Y-Paris | | | | M3Y-Reid | | | |
|---|---|---|---|---|---|---|---|---|
| | $K_0 = 220$ (MeV) | $K_0 = 230$ (MeV) | $K_0 = 240$ (MeV) | $K_0 = 250$ (MeV) | $K_0 = 220$ (MeV) | $K_0 = 230$ (MeV) | $K_0 = 240$ (MeV) | $K_0 = 250$ (MeV) |
| $\rho_t^{Exact}$ (fm$^{-3}$) | 0.091 | 0.093 | 0.094 | 0.095 | 0.090 | 0.092 | 0.093 | 0.094 |
| $\rho_t^{E_{sym}}$ (fm$^{-3}$) | 0.102 | 0.103 | 0.104 | 0.105 | 0.104 | 0.106 | 0.107 | 0.108 |
| $\rho_t^{E_{sym\,4}}$ (fm$^{-3}$) | 0.100 | 0.101 | 0.102 | 0.103 | 0.102 | 0.103 | 0.104 | 0.105 |
| $\rho_t^{E_{sym\,6}}$ (fm$^{-3}$) | 0.097 | 0.098 | 0.099 | 0.100 | 0.098 | 0.099 | 0.100 | 0.102 |
| $\rho_t^{E_{sym\,8}}$ (fm$^{-3}$) | 0.095 | 0.096 | 0.097 | 0.098 | 0.095 | 0.097 | 0.098 | 0.099 |
| $P_t^{Exact}$ (MeV fm$^{-3}$) | 0.485 | 0.505 | 0.511 | 0.513 | 0.551 | 0.573 | 0.581 | 0.589 |
| $P_t^{E_{sym}}$ (MeV fm$^{-3}$) | 0.785 | 0.796 | 0.807 | 0.818 | 0.940 | 0.975 | 0.988 | 1.001 |
| $P_t^{E_{sym\,4}}$ (MeV fm$^{-3}$) | 0.743 | 0.753 | 0.763 | 0.774 | 0.872 | 0.884 | 0.896 | 0.907 |
| $P_t^{E_{sym\,6}}$ (MeV fm$^{-3}$) | 0.707 | 0.717 | 0.728 | 0.738 | 0.812 | 0.824 | 0.856 | 0.867 |
| $P_t^{E_{sym\,8}}$ (MeV fm$^{-3}$) | 0.678 | 0.688 | 0.698 | 0.709 | 0.761 | 0.791 | 0.803 | 0.814 |
| $x_{p(t)}^{Exact}$ | 0.031 | 0.031 | 0.032 | 0.032 | 0.029 | 0.029 | 0.029 | 0.029 |
| $x_{p(t)}^{E_{sym}}$ | 0.028 | 0.029 | 0.029 | 0.029 | 0.028 | 0.028 | 0.028 | 0.028 |
| $x_{p(t)}^{E_{sym4}}$ | 0.029 | 0.029 | 0.029 | 0.029 | 0.028 | 0.028 | 0.028 | 0.028 |
| $x_{p(t)}^{E_{sym6}}$ | 0.030 | 0.030 | 0.030 | 0.030 | 0.028 | 0.029 | 0.029 | 0.029 |
| $x_{p(t)}^{E_{sym8}}$ | 0.030 | 0.030 | 0.030 | 0.030 | 0.028 | 0.029 | 0.029 | 0.029 |